\documentclass{pasj01}
\draft 
\Received{$\langle$17-Jun-2019$\rangle$}
\Accepted{$\langle$24-Jul-2019$\rangle$}
\Published{$\langle$publication date$\rangle$}
\begin{document}

\title{Rotating Ionized Gas Ring around the Galactic Center IRS13E3}
\author{Masato Tsuboi$^{1, 2}$, Yoshimi Kitamura$^1$,  Takahiro Tsutsumi$^3$, Ryosuke Miyawaki$^4$, Makoto Miyoshi$^5$ and Atsushi Miyazaki$^6$ }%
\altaffiltext{1}{Institute of Space and Astronautical Science, Japan Aerospace Exploration Agency,\\
3-1-1 Yoshinodai, Chuo-ku, Sagamihara, Kanagawa 252-5210, Japan }
\email{tsuboi@vsop.isas.jaxa.jp}
\altaffiltext{2}{Department of Astronomy, The University of Tokyo, Bunkyo, Tokyo 113-0033, Japan}
\altaffiltext{3}{National Radio Astronomy Observatory,  Socorro, NM 87801-0387, USA}
\altaffiltext{4}{College of Arts and Sciences, J.F. Oberlin University, Machida, Tokyo 194-0294, Japan}
\altaffiltext{5}{National Astronomical Observatory of Japan, Mitaka, Tokyo 181-8588, Japan}
\altaffiltext{6}{Japan Space Forum, Kanda-surugadai, Chiyoda-ku,Tokyo,101-0062, Japan}

\maketitle
\begin{abstract}
We detected a compact ionized gas associated physically with IRS13E3, an Intermediate Mass Black Hole (IMBH) candidate in the Galactic Center, in the continuum emission at 232 GHz and H30$\alpha$ recombination line using ALMA Cy.5 observation (2017.1.00503.S, P.I.  M.Tsuboi). 
The continuum emission image shows that IRS13E3 is  surrounded by an oval-like structure. The angular size is  $0\farcs093\pm0\farcs006\times 0\farcs061\pm0\farcs004$ ( $1.14\times10^{16}$ cm $\times 0.74\times10^{16}$ cm). 
The structure is also identified in the H30$\alpha$ recombination line. This is seen as an inclined linear feature in the  position-velocity diagram, which is usually  a defining characteristic of  a rotating gas ring around a large mass.
The gas ring has a rotating velocity of $V_\mathrm{rot}\simeq230$ km s$^{-1}$ and an orbit radius of  $r\simeq6\times10^{15}$ cm. From these orbit parameters, the enclosed mass is estimated to be $M_{\mathrm{IMBH}}\simeq2.4\times10^4$ $M_\odot$.  The mass is within the astrometric upper limit mass of the object adjacent to  Sgr A$^{\ast}$. Considering IRS13E3 has an X-ray counterpart, the large enclosed mass would be supporting evidence that IRS13E3 is an IMBH.
Even if a dense cluster corresponds to IRS13E3, the cluster would collapse into an IMBH within  $\tau<10^7$ years due to the very high mass density of $\rho \gtrsim8\times10^{11}  M_\odot pc^{-3}$.
  Because the orbital period is estimated to be as short as $T=2\pi r/V_\mathrm{rot}\sim 50-100$ yr, the morphology of the observed  ionized gas ring is expected to be changed in the next several decades. The mean electron temperature and density of the ionized gas are $\bar{T}_{\mathrm e}=6800\pm700$ K and $\bar{n}_{\mathrm e}=6\times10^5$ cm$^{-3}$, respectively. Then the mass of the ionized gas is estimated to be $M_{\mathrm{gas}}=4\times10^{-4} M_\odot$.
\end{abstract}

\section{Introduction}
Sagittarius A$^\ast$ (Sgr A$^\ast$) is the galactic nucleus of the nearest spiral galaxy, Milky Way, and harbors the Galactic Center black hole (GCBH) with $M\sim4\times10^6 M_\odot$ (e.g. \cite{Boehle}, \cite{Abuter}). This region is a precious laboratory for studying activities and structures of galactic nuclei because peculiar phenomena which will be found in the nuclei of external galaxies by future huge telescopes can be observed by  Atacama Large Millimeter/Submillimeter Array (ALMA) here.
New ones have been discovered in succession since ALMA has started to observe Sgr A$^\ast$ and the surrounding region (e.g. \cite{Moser}, \cite{Tsuboi2017b}, \cite{Yusef-Zadeh2017a}, \cite{Yusef-Zadeh2017b}). 
Although the activity of Sgr A$^\ast$ must be supported by the gravitational energy release of the material accreting to the GCBH,  it is an open question how the GCBH acquires the material up to the present huge mass.
There are two possibilities about the growth history of the GCBH. The first one is that  very dense and massive gas blobs had fallen to  the GCBH. This is a simple scenario but it has several  issues to grow  the GCBH to the present mass.
Even if gas blobs are accreting continuously to the GCBH within the Hubble time, the accretion rate is required to be up to $\dot{M}\sim10^{-4}-10^{-3} ~M_{\odot} ~year^{-1}$. This is several orders of magnitude larger than the present estimated rate  from observations (e.g. \cite{Agol}, \cite{Quataert}). The very high accretion rate requires  a quite low radiation efficiency of $\eta\sim10^{-9}-10^{-8}$ for Sgr A$^\ast$.
 
On the other hand, there is a possibility that the accretion rate may have been much higher in the past caused by any
episodic large accretion event, for example, the formation of the Fermi Bubbles \citep{Su2010} and it
has decreased to the present rate.

The second one is that several heavy compact bodies, stellar-mass black holes and/or intermediate mass black holes (IMBHs), had fallen and merged into the GCBH. 
First of all, it is required that there are such objects in the surrounding region of  Sgr A$^\ast$. 
Recent X-ray observations have already revealed a dozen stellar-mass black holes around Sgr A$^\ast$ \citep{Hailey}.
However, it is an open question in the latter case (Cf. \cite{Schodel2005}, \cite{Paumard}, \cite{Genzel2010}, \cite{Oka}).

In our previous study, we detected a peculiar ionized gas flow associating with the Galactic Center IRS13E complex  which  is located to the southwest of Sgr A$^\ast$ by  $d=4\times10^{17}$ cm ($0.13$ pc)  in projection \citep{Tsuboi2017b}. 
The ionized gas flow has a very large velocity width ($\Delta v_{\mathrm{FWZI}} \sim 650$ km s$^{-1}$) and a very compact size ($r\sim 6\times10^{15}$ cm) in the complex. An extended ionized gas component connecting with the compact component has also been found. 
A hypothesis explains that this uncommon ionized gas system is  a continuous gas flow orbiting  on a high-eccentricity Keplerian orbit around IRS13E3 embedded in  the complex.
The enclosed mass is estimated to be  $10^{4}$ M$_\odot$ by the analysis of the orbit, which is within the mass range of an IMBH. However, the Keplerian orbit parameters derived from the complicated trajectory in the position-velocity diagram should involve large ambiguity. 

If IRS13E3 is an IMBH, an ionized gas ring would be rotating around it. ALMA is able to get the spatial and velocity resolved image of the disk. If the rotating ring is discovered, the enclosed mass derived with the simple structure must be reliable. This would be stronger evidence that the IMBH exists in the complex. We have searched such ionized gas with ALMA in order to prove our hypothesis that IRS13E3 is an IMBH.
Throughout this paper, we adopt 8.0 kpc as the distance to the Galactic center (e.g.\cite{Ghez}, \cite{Gillessen}, \cite{Schodel2009}, \cite{Boehle}): $1\arcsec$ corresponds to  $1.2\times10^{17}$ cm and 0.04 pc at the distance.

\section{Observation and Data Reduction}
We observed the area including Sgr A$^\ast$ and the IRS 13E complex at 230 GHz as an ALMA Cy.5 program (2017.1.00503.S. PI M.Tsuboi).
 The field of view  (FOV) is centered at  $\alpha_{\rm ICRS}$ = $17^{\rm h}45^{\rm m}40^{\rm s}.04$  and $\delta_{\rm ICRS}$= $-29^{\circ}00'28\arcsec.2$, which is a nominal center position of  Sgr A$^\ast$. The diameter of the FOV is $\sim25\arcsec$ in FWHM. The center frequencies of the spectral windows are  $f_c=217.5, 219.5, 234.0,$ and $231.9$ GHz, respectively. 
The last spectral window covers the H30$\alpha$ recombination line (231.9 GHz). The frequency coverage and frequency resolution are 2.0 GHz and 15.625 MHz, respectively.   The velocity resolution corresponds to a velocity width of $\sim20$ km s$^{-1}$, which approximately corresponds to the thermal velocity width of ionized gas at $1\times10^4$ K.
 The observations were performed in ten days from 5 Oct. 2017 to 20 Oct. 2017. 
The observation epochs were in the period of the longest baseline antenna configuration of ALMA.
J1744-3116 and J1752-2956 were used as a phase calibrator. The flux density scale was determined using J1924-2914. 

We performed the data analysis by Common Astronomy Software Applications (CASA) \citep{McMullin}. 
Before imaging in the recombination line, the continuum emissions of Sgr A$^\ast$ and the Galactic Center Mini Spiral (GCMS) were subtracted from the data using CASA task, {\tt uvcontsub (fitorder=1)} for the combined data in the ten-day observations. The imaging to obtain channel maps was done using CASA 5.4 with {\tt clean} task. We also got the line-free continuum data using this task. This data was used for the continuum imaging shown below. We will show the full results including the resultant continuum maps of the other three spectral windows in another publication. 
The angular resolution using ``natural weighting" is $0\arcsec.037 \times 0\arcsec.024, PA=87^\circ$.  The sensitivities of the original velocity width channels are $1\sigma=90~ \mu$Jy beam$^{-1}$ or  2.3 K in $T_\mathrm{B}$ in the emission-free areas. 
The sensitivity is slightly worse than the expected one by the ALMA sensitivity calculator because there are yet unavoidable sidelobes of Sgr A$^\ast$ in the channel maps. 

The flux density of Sgr A$^\ast$ changed significantly in the observation epochs although such variability had been reported (e.g. \cite{Miyazaki}).  The flux density in the first seven days (from 5 Oct. 2017 to 14 Oct. 2017) was around $S_\nu(\mathrm{Sgr A}^{\ast}) \sim2.4$ Jy. On the other hand, the flux density in the last three days (from 16 Oct. 2017 to 20 Oct. 2017) was around  $S_\nu(\mathrm{Sgr A}^{\ast}) \sim1.1$ Jy.  Such significant flux density change might make some artifacts including unavoidable sidelobe effects in the  continuum map from the combined data. We used only the data of the first seven days to produce the resultant continuum map here.  In addition, the data with the baselines over 11 km is removed because this seems to increase the sidelobe effects in the continuum map.The complex gain errors of the data were minimized using the ``self-calibration" method in CASA.The  ``self-calibration" method give us to obtain a high dynamic range in the continuum map. The angular resolution using ``natural weighting" is $0\arcsec.037 \times 0\arcsec.025, PA=85^\circ$, which is almost the same as that of the channel maps. The sensitivity is $1\sigma=50~ \mu$Jy beam$^{-1}$ or  1.3 K in $T_\mathrm{B}$ in the emission-free areas. Although the sidelobe effects still remain around Sgr A$^\ast$, the dynamic range reaches to  $>40000$ in the other areas  of the map. 

\begin{figure}
\begin{center}
\includegraphics[width=18cm, bb=0 0 994.4 1081.97]{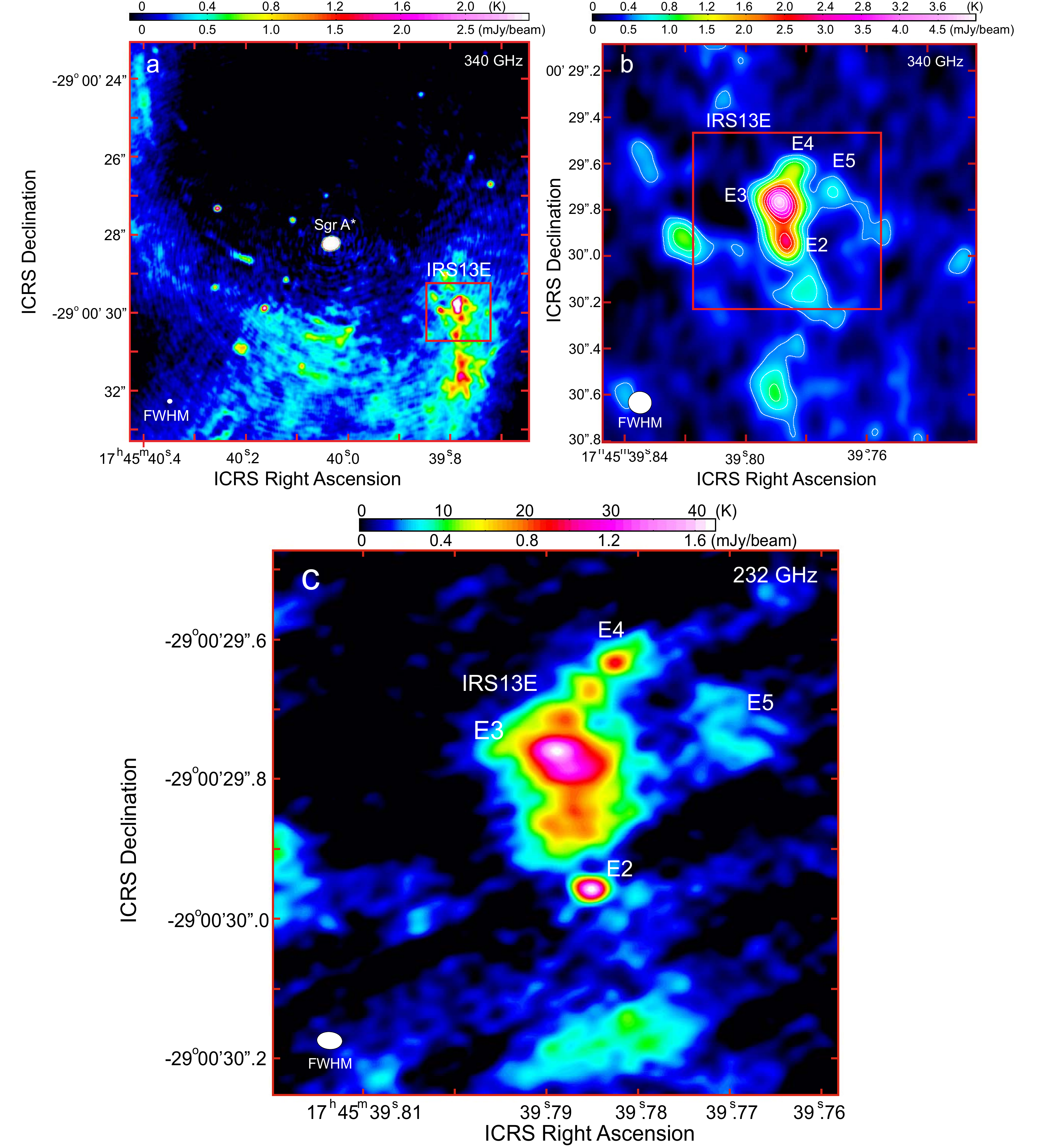}
 \end{center}
 \caption{{\bf a} Continuum map at 340 GHz of the Sgr A$^\ast$ region \citep{Tsuboi2017b}.  The angular resolution is $0\arcsec.12 \times 0\arcsec.11, PA=-84^\circ$  in FWHM, which is shown as the oval at the lower left corner.  {\bf b} Close-up view of the IRS13E complex at 340 GHz. The map area shows the rectangle in {\bf a}. These are finding charts of the Galactic Center IRS13E complex. {\bf c} Continuum map at 232 GHz of the IRS13E complex. The map area shows the rectangle in {\bf b}. The angular resolution is $0\arcsec.037 \times 0\arcsec.025, PA=85^\circ$  in FWHM, which is shown as the oval at the lower left corner. The peak position at 232 GHz of IRS13E3 is $\alpha_{\rm ICRS}=17^{\rm h}45^{\rm m}39^{\rm s}.789$,  $\delta_{\rm ICRS}=-29^\circ00\arcmin29\arcsec.759$. The peak flux density is $S_\nu=1.712\pm0.054$ mJy beam$^{-1}$. 
}
 \label{Fig1}
\end{figure}

\begin{figure}
\begin{center}
\includegraphics[width=18cm, bb=0 0 2233.53 1633.88]{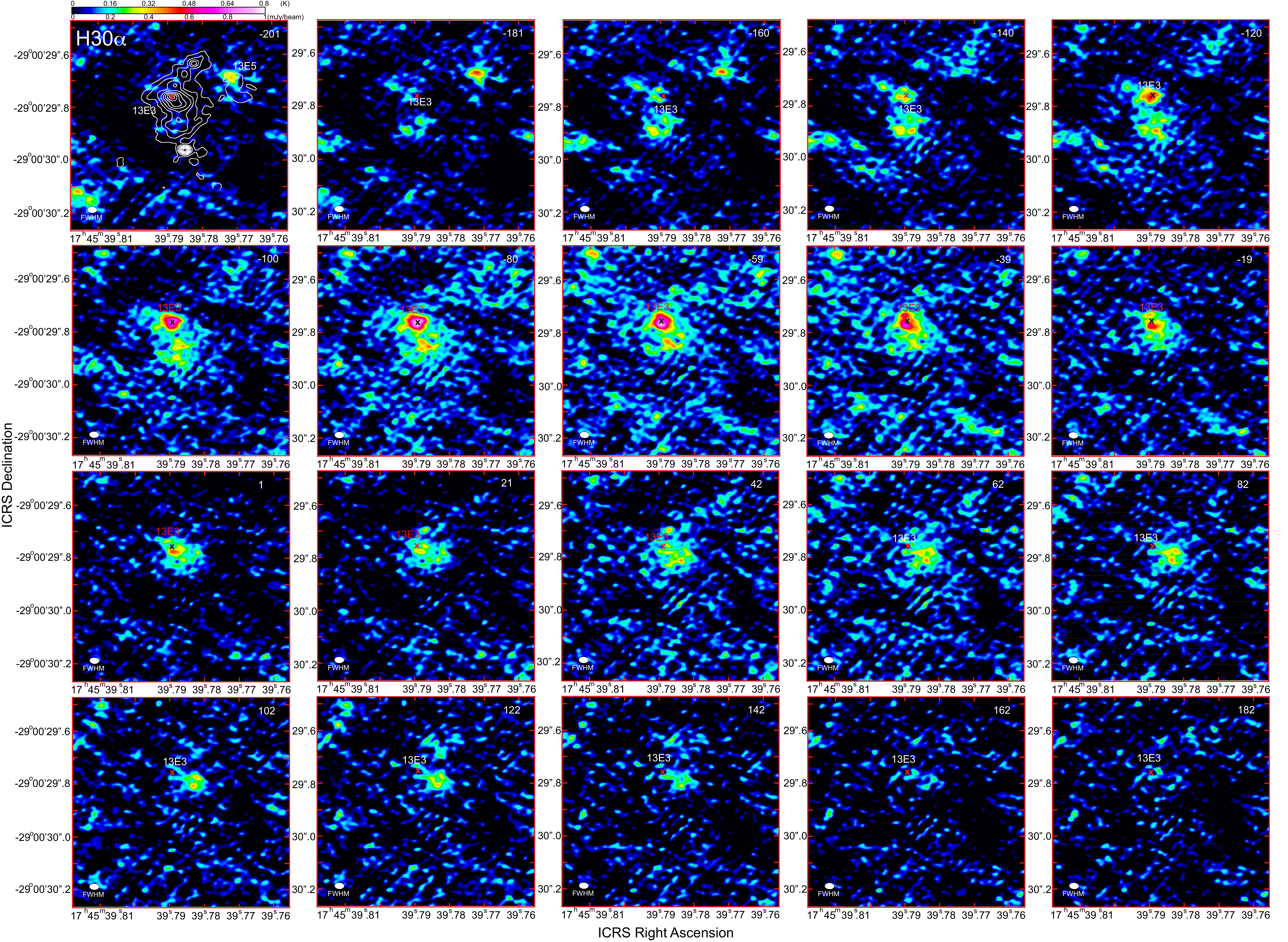}
 \end{center}
 \caption{ Channel maps of the Galactic Center IRS13E complex in the H30$\alpha$ recombination line. The central velocity ranges from $V_{\mathrm{c, LSR}}=-201$ to $+182$ km s$^{-1}$, and the velocity width of each panel is $\sim20$ km s$^{-1}$.  The angular resolution is $0\arcsec.037 \times 0\arcsec.024, PA=87^\circ$ in FWHM, which is shown as the oval at the lower left corner of each panel.  
Contours in the upper-right panel show the continuum emission of IRS13E3 at 232 GHz for comparison (see Figure 1c).   The first contour and interval are both 5 K in $T_\mathrm{B}$.  The crosses in each panel show the peak position of the continuum emission. }
 \label{Fig2}
\end{figure}

\section{Results}
Figure 1a  shows the continuum map of the vicinity of Sgr A$^\ast$  at 340 GHz \citep{Tsuboi2017b}.  Figure 1b shows the close-up view of the GC IRS13E complex at 340 GHz, of which area is shown as the rectangle in Figure 1a. These are finding charts of   IRS13E3 in the complex. The angular resolution is $0\arcsec.12 \times 0\arcsec.11, PA=-84^\circ$  in FWHM.
The complex is resolved into a group of compact objects in the maps. Most of these objects are the IR objects  identified in the previous observations (e.g. \cite{Maillard}, \cite{Schodel2005}, \cite{Paumard}). The strongest component is IRS13E3.
Figure 1c shows the continuum map at 232 GHz of the IRS13E complex by this observation. The map area is shown as the rectangle in Figure 1b. The angular resolution is $0\arcsec.037 \times 0\arcsec.025, PA=85^\circ$  in FWHM, which is shown as the oval at the lower left corner.    IRS13E3 is clearly resolved into an inclined oval-like structure which is elongated in the direction of northeast to southwest.  
The angular source size of the structure is derived to be $\theta_{\mathrm{maj. obs.}}\times\theta_{\mathrm{min. obs.}}= 0\farcs093\pm0\farcs006\times 0\farcs061\pm0\farcs004$, $PA\sim64^\circ$  by the two-dimensional Gaussian fit including beamsize correction. These correspond to $d_{\mathrm{maj. obs.}}\times d_{\mathrm{min. obs.}}= $ 
 $1.14\times10^{16}$ cm $\times 0.74\times10^{16}$ cm
at the the  Galactic center distance. The total flux density of the oval-like structure is $S_\nu=10.52\pm0.90$ mJy.
 The flat spectrum in mm wavelength suggests that the emission is an optically thin free-free emission \citep{Tsuboi2017b}.  
Moreover,  finer structures are identified in the oval-like structure. The intensity peak  of IRS13E3 at 232 GHz is located around the northeast side of the oval-like structure. The position is $\alpha_{\rm ICRS}=17^{\rm h}45^{\rm m}39^{\rm s}.7888(\pm0^{\rm s}.00005)$,  $\delta_{\rm ICRS}=-29^\circ00\arcmin29\arcsec.7614(\pm0\arcsec.0003)$.  The peak flux density is $S_\nu=1.712\pm0.054$ mJy beam$^{-1}$.  

A ridge-like structure extending  from south to north crossing the oval-like structure is also identified in Figure 1c. This is probably a counter part of the ionized gas  flow orbiting on a Keplerian orbit around IRS13E3, which  has been identified in the previous observation \citep{Tsuboi2017b}.  The major axis of the oval-like structure is not parallel to the central axis of the extended gas flow and the crossing angle is about $45^\circ$. 
The total flux density of the extended gas flow is $S_\nu=5.80\pm0.59$ mJy. 
There are also the components corresponding to the famous Wolf-Rayet stars, IRS13E2 and IRS13E4, which have no spread larger than the beam size. 

Figure 2 shows the channel map of the vicinity of IRS13E3 in the H30$\alpha$ recombination line with the central velocities ranging from  $V_{\mathrm{c, LSR}}=-201$ to $+182$ km s$^{-1}$. The map area is the same as Figure 1c. The angular resolution of these maps is $0\farcs037 \times 0\farcs024, PA=87^\circ$ in FWHM. The velocity width of the channel maps is $\Delta v= 20.2$ km s$^{-1}$.  The contours in Figure 2 (the upper-right panel)  shows the continuum emission of IRS13E3 at 232 GHz for comparison (see Figure 1c). Ionized gas components corresponding to the oval-like structure of IRS13E3 are identified in these maps. The components positionally shift from northeast to southwest along the major axis of the oval-like structure with increasing velocity. The component with negative velocity of $V_{\mathrm{c, LSR}}= 181$ to 1 km s$^{-1}$ is brighter than that with positive velocity of $V_{\mathrm{c, LSR}}= 21$ to 182 km s$^{-1}$. 

Figure 3 shows the integrated intensity maps  with the velocity ranges of  $V_{\mathrm{LSR}}=-190$ to $10$ km s$^{-1}$ (left) and $V_{\mathrm{LSR}}=10$ to $210$ km s$^{-1}$ (right). These show the components with negative and positive velocities mentioned above, respectively.
The angular resolution of these maps is the same as that of  Figure 2. The contours in Figure 3 also shows the continuum emission at 232 GHz for comparison.  The whole distribution of the ionized gas is oval-like but biased on the western side of the continuum peak. 
 As seen in Figure 3a, the peak positionin the H30$\alpha$ recombination line is $\alpha_{\rm ICRS}=17^{\rm h}45^{\rm m}39^{\rm s}.7896(\pm0^{\rm s}.0001)$,  $\delta_{\rm ICRS}=-29^\circ00\arcmin29\arcsec.7704(\pm0\arcsec.0005)$. The integrated intensity peak position of the H30$\alpha$ recombination line is slightly shifted from that of the continuum emission.  
 
Another component, which is located south  of the oval-like structure, is identified in the channel maps with $V_{\mathrm{c, LSR}}=-181$ to $-39$ km s$^{-1}$.  This component positionally shift from south to north  with increasing velocity.
The component is also identified in the integrated intensity map (see Figure 3a).  This is a counter part of the component extending to the south seen in the continuum map, Figure 1c.  
Many faint components are widely distributed in the panels with $V_{\mathrm{c, LSR}}=-80$, $-59$, and $-39$ km s$^{-1}$.  These would belong to ``Bar" of the GCMS which is extended widely (e.g. Figure 3 in \cite{Tsuboi2017b}). The major part of ``Bar" is probably resolved out at the high resolution in this observation. In addition, the extended component corresponding to IRS13E5 is also identified in the panels with $V_{\mathrm{c, LSR}}=-201$ to $-140$ km s$^{-1}$. However, the components corresponding to IRS13E2 and IRS13E4 are not clearly found.

\begin{figure}
\begin{center}
\includegraphics[width=18cm, bb=0 0 981.79 431.15]{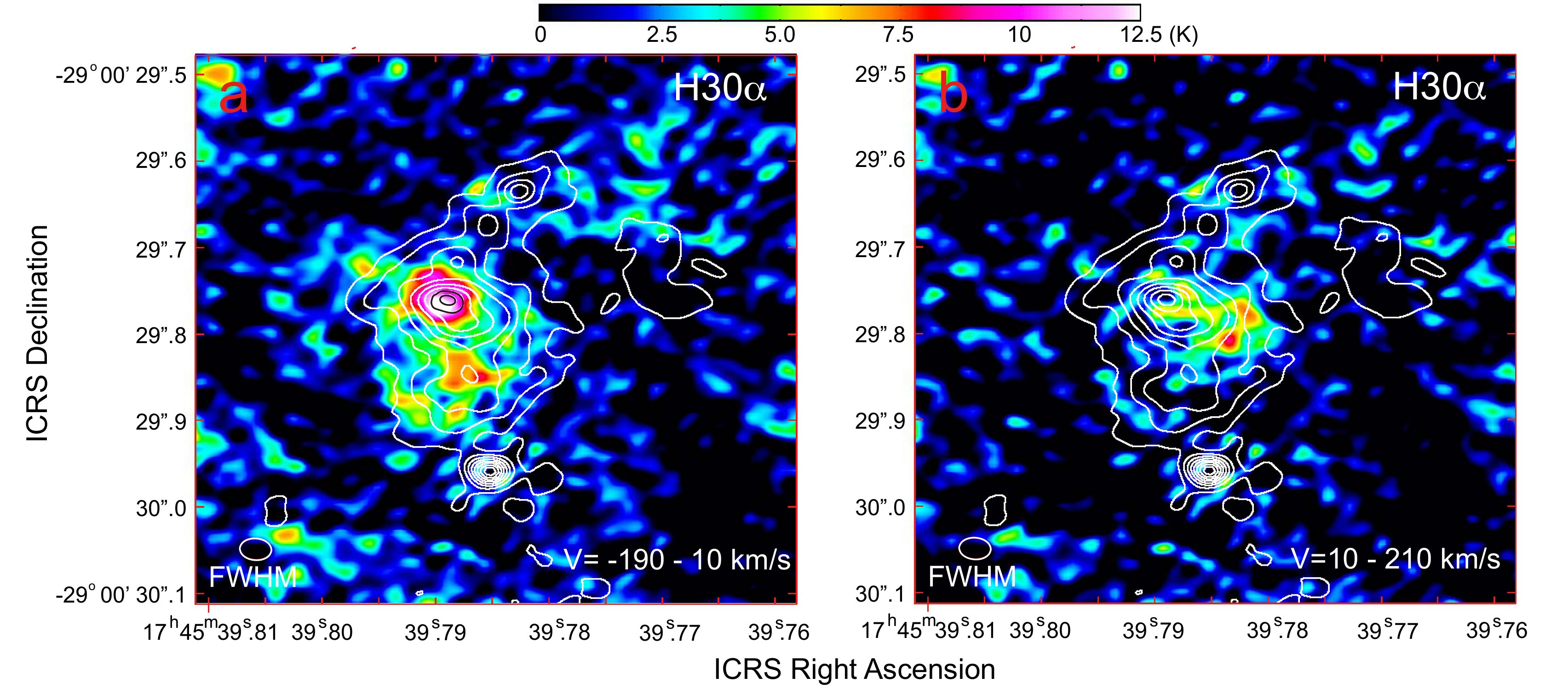}
 \end{center}
 \caption{Integrated intensity maps  in the H30$\alpha$ recombination line  (231.9 GHz) with the velocity ranges of  {\bf a} $V_{\mathrm{LSR}}=-190$ to $10$ km s$^{-1}$  and  {\bf b} $V_{\mathrm{LSR}}=10$ to $210$ km s$^{-1}$. The angular resolution of these maps is $0\farcs037 \times 0\farcs024, PA=87^\circ$, which is shown as the oval at the lower left corner of each panel. Contours show the continuum emission of IRS13E3 at 232 GHz for comparison.  The first contour and interval are both 5 K in $T_\mathrm{B}$. The crosses in the panels show the peak position, $\alpha_{\rm ICRS}=17^{\rm h}45^{\rm m}39^{\rm s}.789,  \delta_{\rm ICRS}=-29^\circ00\arcmin29\arcsec.761$.}
 \label{Fig3}
\end{figure}
\begin{figure}
\begin{center}
\includegraphics[width=18cm, bb=0 0 1067.15 865.98]{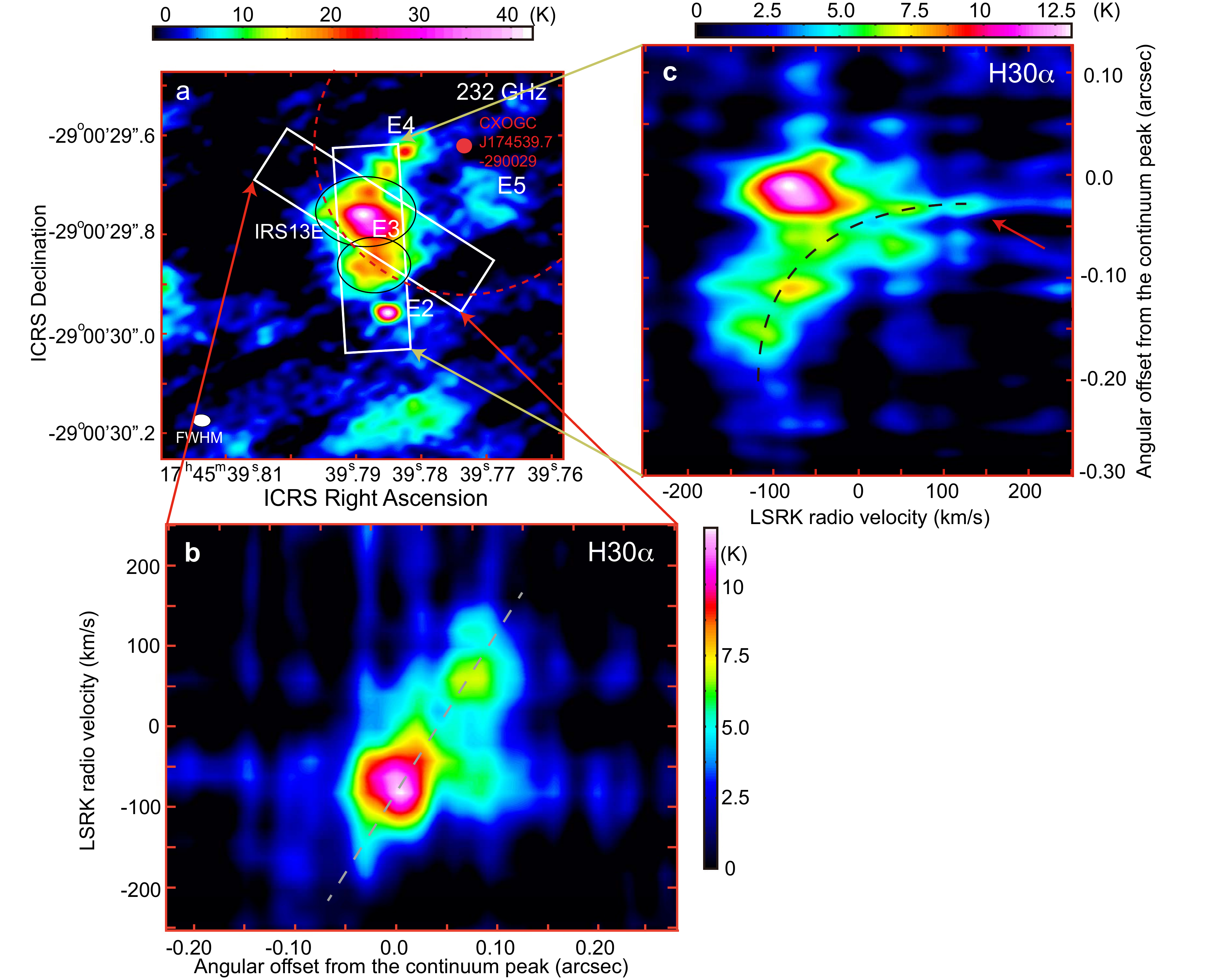}
 \end{center}
 \caption{{\bf a} Continuum map at 232 GHz which is the same as shown in figure 1c. The total flux density at 232 GHz of the ionized gas ring around IRS13E3 and the extended gas flow are $S_\nu=10.52\pm0.90$ mJy and  $S_\nu=5.80\pm0.59$ mJy, respectively. The integration areas are shown  as ovals.
 The position of CXOGC174539.7-290029 ($\alpha_{\rm ICRS}=17^{\rm h}45^{\rm m}39^{\rm s}.773$,  $\delta_{\rm ICRS}=-29^\circ00\arcmin29\arcsec.622$) is shown as a red filled circle. The broken red circle is the $95\%$ position error circle which has the radius of $0\arcsec.3$.
 {\bf b} Position-velocity (PV) diagram along the major axis of the inclined oval-like  ionized gas around IRS13E3. The integration area is shown as a white rectangle in {\bf a} .  {\bf c} PV diagram along  the ionized gas  flow orbiting on a Keplerian orbit around the IRS13E3 (curve broken line; \cite{Tsuboi2017b}).  The integration area is shown as a white rectangle in {\bf a}. }
 \label{Fig4}
\end{figure}

\section{Discussion}
\subsection{Enclosed Mass of IRS13E3}
Figure 4b shows the position-velocity (PV) diagram in the H30$\alpha$ recombination line along the major axis of the oval-like structure of IRS13E3. The integration area is shown as the red rectangle in Figure 4a which is the continuum map at 232 GHz shown in figure 1c. 
The oval-like structure is seen as an inclined linear feature in Figure 4b. Generally, such a feature in a PV map indicates a ring like structure rotating around a massive object.  The observed feature may be a part of an ionized gas ring rotating around  IRS13E3.  
The rotating velocity of the structure is derived to be $V_\mathrm{rot}\sim150/\cos{i}$  km/s from the difference between the positive and negative ends of the linear feature after correction of thermal broadening. Here $i$ is the inclination angle of the structure ($i=0^\circ$ for edge-on). 
The inclination angle is derived to be $i \sim 50^\circ$ from the aspect ratio of the Gaussian-fit angular size mentioned in Results.
Then the rotating velocity is estimated to be $V_\mathrm {rot} \sim 230$ km s$^{-1}$.
While the radius of the orbit is derived to be $r\sim6\times10^{15}$ cm from the semi-major axis of the angular size.  
Assuming that the ionized gas obeys a single circular orbit simply, the enclosed mass is estimated to be $M\sim2.4\times10^4$  $M_\odot$  from the above parameters using the formula; $M=\frac{rV_\mathrm {rot}^2}{G}$.
Although the mass is smaller than the enclosed mass based on the analysis for the Keplerian orbit of the larger extended gas flow \citep{Tsuboi2017b},  it is still in the mass range for IMBH.  This mass is within the upper limit mass of the object adjacent to  Sgr A$^{\ast}$, which had been derived from the astrometry of Sgr A$^{\ast}$ using VLBA \citep{Reid}. 
Of course, massive compact objects do not uniquely imply BHs and the alternate may be a compact cluster of star or star remnant and so on.  From the derived mass and orbit radius, the lower limit of the mass density is estimated to be $\rho \gtrsim8\times10^{11}  M_\odot pc^{-3}$.
 According to the previous estimations (e.g. \cite{Maoz}), the cluster with such mass density and such mass would be collapsed within less than $\tau<10^7$ years and form a BH eventually.
Moreover, IRS13E3 is located within the $95\%$ error circle ($0\arcsec.3$) of the X-ray point source detected by CHANDRA; CXOGCJ174539.7-290029 \citep{Muno}.  The position and the error circle are shown in Figure 4a.
The X-ray flux at $2-8$ keV of the source is reported to  $F_\mathrm {2-8}=1.12\times10^{-5}$ ph cm$^{-2}$ s$^{-1}$, which is as large as about half that of SgrA$^{\ast}$,  $F_\mathrm {2-8}=2.44\times10^{-5}$ ph cm$^{-2}$ s$^{-1}$. 
If a modest amount of gas accretes onto an IMBH, strong X-ray emission is expected to be produced. 
IRS13E3 has such a X-ray counterpart.
Therefore this enclosed mass would be the convincing evidence that the IMBH exists in the complex.

On the other hand, figure 4c shows the PV diagram along the the south-extended gas flow.  The integration area is shown as the white rectangle in figure 4a. The gas flow is seen as a curved feature in the PV diagram (a broken line curve), which is clearly distinguished from the gas ring mentioned above. 
This gas flow may correspond to a part of the Keplerian orbit around IRS13E3 identified by our previous observation \citep{Tsuboi2017b}. 
There is a faint component $\sim0\arcsec.03$ from the continuum peak  in the PV diagram (an arrow), which is the positive velocity end of  the curved feature. The physical distance is $R\sim 3.7\times10^{15}$ cm. The radial velocity of the component is $V_\mathrm {R} \sim 150$ km s$^{-1}$. 
Although, the north extension of the major axis of the gas flow does not coincide with the continuum peak position of IRS13E3 exactly (see figure 1c), we assume here that the gas flow is now falling toward IRS13E3 freely. 
We suppose that the gas flow had been nearly at rest  in the far distance and the rest radial velocity was $V_\mathrm {\infty} \sim -120$ km s$^{-1}$ from Figure 4c. Thus the gain in velocity during falling is $\Delta V=V_\mathrm {R}-V_\mathrm {\infty}\sim 270$ km s$^{-1}$.
The mass is estimated to be $M\sim1.0\times10^4/(\cos{j})^2M_\odot$   using the formula; $M=\frac{R\Delta V^2}{2G}$. Here $j$ is the angle between the moving direction of the gas flow and the line-of-sight ($j=0^\circ$ for the accord of the directions).  This is roughly consistent with the enclosed mass mentioned above. The observed continuum image of IRS13E3 has inner structures  as mentioned in the previous section (see figure 1c). 
These should show that the ionized gas does not occupy the orbit homogeneously. Because the orbital period is estimated to be $T=2\pi r/V_\mathrm {rot}\sim 50-100$ yr, the morphology of the observed ionized gas is expected to change significantly in the next several decades.

\subsection{Physical Properties of the Rotating Ionized Gas Ring around IRS13E3}
The LTE electron temperature, $T^\ast_{\mathrm e}$, of the ionized gas ring around IRS13E3 is estimated from the ratio between the integrated recombination line intensity, $\int S_{\mathrm{line}}(\mathrm{H}30\alpha)dV $, and the continuum flux density, $S_{\mathrm{cont}}(\mathrm{232GHz})$, assuming that the line and continuum emissions are optically thin. 
The well-known formula of the LTE electron temperature is given by
\begin{equation}
\label{1}
T^\ast_{\mathrm e}[\mathrm K]=\left[\frac{6.985\times10^3}{a(\nu, T^\ast_{\mathrm e})}\Big(\frac{\nu}{\mathrm{GHz}}\Big)^{1.1}
\frac{1}{1+\frac{N(\mathrm{He^+})}{N(\mathrm{H^+})}}
\frac{S_{\mathrm{cont}}(\nu)}
{\int S_{\mathrm{line}}\Big(\frac{dV}{\mathrm{km~s}^{-1}}\Big)}
\right]^{\frac{1}{1.15}}.
\end{equation}
The correction factor, $a(\nu, T^\ast_{\mathrm e})$, at $\nu=232$ GHz is calculated to be $a=0.763-0.892$  for $T^\ast_{\mathrm e}=0.5-1.5\times10^4$ K \citep{Mezger}.  We assume that the number ratio of He$^+$ to H$^+$ is $\frac{N(\mathrm{He^+})}{N(\mathrm{H^+})}=0.09$, a typical value in the Galactic center region (e.g. \cite{Tsuboi2017}). The LTE electron temperature is obtained by iteratively solving the formula for $T^\ast_{\mathrm e}$. 
The total flux densities at 232 GHz of the ionized gas ring around IRS13E3 and the extended gas flow are $S_\nu=10.52\pm0.90$ mJy and  $S_\nu=5.80\pm0.59$ mJy, respectively. 
The integration areas are shown in Figure 4a as the two ovals.
Meanwhile the total integrated intensity of the H30$\alpha$ recombination line emission in these areas are $\int S_\nu dv=1.31\pm0.13$ Jy km s$^{-1}$ and $\int S_\nu dv=0.68\pm0.07$ Jy km s$^{-1}$, respectively. 
The mean electron temperatures of the ionized gas ring and the extended gas flow are estimated to be $\bar{T}_{\mathrm e}=6800\pm700$ K and $\bar{T}_{\mathrm e}=7100\pm700$ K, respectively. 

Using the observed brightness temperature of the continuum emission and the derived electron temperature, the electron density, $n^\ast_{\mathrm e}$, of the ionized gas ring around IRS13E3 is estimated by the following well-known formula;
\begin{equation}
\label{2}
n_{\mathrm e}[{\mathrm cm}^{-3}]=\left[\frac{T_{\mathrm B}T_{\mathrm e}^{\ast 0.35}\Big(\frac{\nu}{\mathrm{GHz}}\Big)^{2.1}}{2.674\times10^{-20}\alpha(\nu,T)\Big(\frac{L}{\mathrm{cm}}\Big)}\right]^{0.5}
\end{equation}
\citep{Altenhoff}.
The mean electron density is $\bar{n}_{\mathrm e}=6\times10^5$ cm$^{-3}$ assuming that the ionized gas in IRS13E3 has a spherical shape with a diameter of  $ d\sim1.14\times10^{16}$ cm or the path length is  $ L\sim1.14\times10^{16}$ cm (see Results). The mass of the ionized gas ring is $M=\frac{4\pi}{3}\big(\frac{L}{2}\big)^3\bar{n}_{\mathrm e}m_{\mathrm{H}}=4\times10^{-4} M_\odot$. 

\section{Conclusions}
We obtained the spatially resolved images of the compact ionized gas associated with IRS13E3 in the continuum emission at 232 GHz and H30$\alpha$ recombination line using ALMA,  which is an IMBH candidate in the Galactic Center. 
The continuum emission image shows that IRS13E3 is  surrounded by an oval-like structure. The angular size of the structure is  $0\farcs093\pm0\farcs006\times 0\farcs061\pm0\farcs004$( $1.14\times10^{16}$ cm $\times 0.74\times10^{16}$ cm). The structure is seen as an inclined linear feature in the H30$\alpha$ position-velocity diagram.
We consider that the ionized gas is a ring-like structure rotating around IRS13E3 with the rotating velocity of $V_\mathrm{rot}\simeq230$ km s$^{-1}$ and the orbit radius of $r\simeq6\times10^{15}$ cm.
From these orbit parameters, the enclosed mass is estimated to be $M\simeq2.4\times10^4$  $M_\odot$. 
The mass is within the astrometric upper limit mass of the object adjacent to  Sgr A$^{\ast}$. Considering IRS13E3 has an X-ray counterpart, the large enclosed mass would be supporting evidence that IRS13E3 is an IMBH.
Even if a dense cluster corresponds to IRS13E3, the cluster would collapse into an IMBH within  $\tau<10^7$ years due to the very high mass density of $\rho \gtrsim8\times10^{11}  M_\odot pc^{-3}$. 
Because the orbital period is estimated to be $T=2\pi r/V_\mathrm{rot}\sim 50-100$ yr, the morphology of the observed structure is expected to change in the next several decades. The mean electron temperature and density of the ionized gas ring are $\bar{T}_{\mathrm e}=6800\pm700$ K and $\bar{n}_{\mathrm e}=6\times10^5$ cm$^{-3}$, respectively. Then the mass of the ionized gas ring is $M=4\times10^{-4} M_\odot$. 

\begin{ack} 
This work is supported in part by the Grant-in-Aid from the Ministry of Eduction, Sports, Science and Technology (MEXT) of Japan, No.19K03939. The National Radio Astronomy Observatory (NRAO) is a facility of the National Science Foundation operated under cooperative agreement by Associated Universities, Inc. USA.  ALMA is a partnership of ESO (representing its member states), NSF (USA) and NINS (Japan), together with NRC (Canada), NSC and ASIAA (Taiwan), and KASI (Republic of Korea), in cooperation with the Republic of Chile. The Joint ALMA Observatory (JAO) is operated by ESO, AUI/NRAO and NAOJ. This paper makes use of the following ALMA data: ADS/JAO.ALMA\#2015.1. 01080.S and ALMA\#2017.1.00503.S.  

\end{ack}

\clearpage

\end{document}